# How to calculate the practical significance of citation impact differences? An empirical example from evaluative institutional bibliometrics using adjusted predictions and marginal effects

Lutz Bornmann* and Richard Williams**

*Division for Science and Innovation Studies

Administrative Headquarters of the Max Planck Society

Hofgartenstr. 8,

80539 Munich, Germany.

E-mail: bornmann@gv.mpg.de

**Department of Sociology

810 Flanner Hall

University of Notre Dame

Notre Dame, IN 46556 USA

E-mail: Richard.A.Williams.5@ND.Edu

Web Page: http://www.nd.edu/~rwilliam/

**Abstract**

Evaluative bibliometrics is concerned with comparing research units by using statistical procedures. According to Williams (2012) an empirical study should be concerned with the substantive and practical significance of the findings as well as the sign and statistical significance of effects. In this study we will explain what adjusted predictions and marginal effects are and how useful they are for institutional evaluative bibliometrics. As an illustration, we will calculate a regression model using publications (and citation data) produced by four universities in German-speaking countries from 1980 to 2010. We will show how these predictions and effects can be estimated and plotted, and how this makes it far easier to get a practical feel for the substantive meaning of results in evaluative bibliometric studies. We will focus particularly on Average Adjusted Predictions (AAPs), Average Marginal Effects (AMEs), Adjusted Predictions at Representative Values (APRVs) and Marginal Effects at Representative Values (MERVs).

# 1     Introduction

Evaluative bibliometrics is concerned with comparing research units: Has Researcher 1 performed better during his or her career so far than Researcher 2? Has University 1 achieved a higher citation impact over the last five years than University 2? Good examples of comparative evaluations are the Leiden Ranking 2011/2012 (Waltman et al., 2012) and the SCImago Institutions Ranking (SCImago Reseach Group, 2012), in which different bibliometric indicators are used to compare higher education institutions and research-focused institutions. As well as assessing the research output (measured by the number of publications), the evaluations measure primarily the citation impact, an important aspect of research quality. If sophisticated methods are employed in the evaluation, field and age normalised indicators are used to measure the citation impact. We consider $PP_{top\ 10\%}$ to be currently the bibliometric indicator which should be preferred in the evaluation of institutions. $PP_{top\ 10\%}$ is the proportion of an institution's publications which belong to the top 10% most frequently cited publications; a publication belongs to the top 10% most frequently cited if it is cited more frequently than 90% of the publications published in the same field and in the same year. $PP_{top\ 10\%}$ is seen as the most important indicator in the Leiden Ranking by the Centre for Science and Technology Studies (Leiden University, The Netherlands): "We therefore regard the $PP_{top\ 10\%}$ indicator as the most important impact indicator in the Leiden Ranking" (Waltman, et al., 2012).

For an evaluation study, a population, defined as the whole bibliometric data for an institution, is usually split up into natural, non-overlapping groups such as different publication years (Bornmann & Mutz, 2013). Such groups provide for clusters in a two-stage sampling design ("cluster sampling"), in which, firstly, one single cluster is randomly selected from a set of clusters (Levy & Lemeshow, 2008). For example, for an evaluation study, the clusters would consist of ten consecutive publication years (e.g. cluster 1: 1971 to 1980;

cluster 2: 1981 to 1990 …). Secondly, all the bibliometric data (publications and corresponding metrics) is gathered (census) for the selected cluster (e.g. cluster 2). Waltman, et al. (2012) include the 2005-2009 cluster in the Leiden Ranking 2011/2012 mentioned above. With statistical tests it is possible to verify the statistical significance of results (such as performance differences between two universities) on the basis of a cluster sample. If a statistical test which looks at the difference between two institutions with regard to their performances turns out to be statistically significant, it can be assumed that the difference has not arisen by chance, but can be interpreted beyond the data at hand (the results can be related to the population).

According to Williams (2012) a study should be concerned with the substantive and practical significance of the findings as well as the sign and statistical significance of effects. Unfortunately, for many techniques, such as logistic regression, the practical significance of a finding may be difficult to determine from the model coefficients alone. For example, if the coefficient for X1 is .7, we may be able to easily determine that the effect of X1 is positive and statistically significant. But, it is much harder to tell whether those with higher scores on X1 are slightly more likely to experience an event, moderately more likely, or much more likely. Further complicating things is that, in logistic regression, the effect that increases in X1 will have on the probability of an event occurring will vary with the values of the other variables in the model. For example, Williams (2012) shows that the effect of race on the likelihood of having diabetes is very small at young ages, but steadily increases at older ages.

Hence, as Long and Freese (2006) show, results can often be made more tangible by computing predicted/expected values for hypothetical or prototypical cases. For example, if we want to get a practical feel for the performance differences between two universities in a logistic regression model, we might compare the predicted probabilities of $P_{top\ 10\%}$ for two publications (from the different universities) which both have low, average, and/or high values for other variables in the model which might have an effect on citation impact (e.g.

publication in low versus high impact journals). Such predictions are sometimes referred to as margins, predictive margins, or (our preferred terminology) adjusted predictions. Another useful aid to interpretation are marginal effects, which can, for example, show succinctly how the adjusted predictions for university 1 differ from the adjusted predictions for university 2.

In this study we will explain what adjusted predictions and marginal effects are and how useful they are for institutional evaluative bibliometrics. As an illustration, we will calculate a regression model using publication and citation data for four universities (univ1, univ 2, univ 3, and univ 4). We will show how these predictions and effects can be estimated and plotted, and how this makes it far easier to get a practical feel for the substantive meaning of results in evaluative bibliometric studies. We will focus particularly on Average Adjusted Predictions (AAPs), Average Marginal Effects (AMEs), Adjusted Predictions at Representative Values (APRVs) and Marginal Effects at Representative Values (MERVs).

## 2 Methods

### 2.1 Description of the data set and the variables

Publications produced by four universities in German-speaking countries from 1980 to 2010 are used as data (see Table 1). The data was obtained from InCites (Thomson Reuters). InCites (http://incites.thomsonreuters.com/) is a web-based research evaluation tool allowing assessment of the productivity and citation impact of institutions. The metrics (such as the percentiles for each individual publication) are generated from a dataset of 22 million Web of Science (WoS, Thomson Reuters) publications from 1980 to 2010. The calculation of $PP_{top\ 10\%}$ or the determination of the top 10% most cited publications ($P_{top\ 10\%}$) is based on percentile data.

*Table 1 about here*

Percentiles are defined by Thomson Reuters as follows: "The percentile in which the paper ranks in its category and database year, based on total citations received by the paper.

The higher the number [of] citations, the smaller the percentile number [is]. The maximum percentile value is 100, indicating 0 cites received. Only article types *article, note, and review* are used to determine the percentile distribution, and only those same article types eceive a percentile value. If a journal is classified into more than one subject area, the percentile is based on the subject area in which the paper performs best, i.e. the lowest value" http://incites.isiknowledge.com/common/help/h_glossary.html). Since in a departure from convention low percentile values mean high citation impact (and vice versa), the percentiles received from InCites are called "inverted percentiles." To identify $P_{top\ 10\%}$, publications from the universities with an inverted percentile smaller than or equal to 10 are coded as 1; publications with an inverted percentile greater than 10 are coded as 0.

As Table 1 shows, $PP_{top\ 10\%}$ for all the universities is 20.7%. The universities thus have a 10.7% higher $PP_{top\ 10\%}$ than one could expect were one to put together a sample consisting of percentiles for publications randomly in InCites (the expected value is therefore 10). As the distribution of publications over the universities in Table 1 shows, there are many more publications for univ 3 and univ 4 than for univ 1 and univ 2. In addition to the universities, other independent variables which have been shown in other studies to influence the citation impact of publications have been included in the regression model (see the overview in Bornmann & Daniel, 2008): (1) The more authors a publication has and the longer it is, the greater its citation impact. (2) According to Bornmann, Mutz, Marx, Schier, and Daniel (2011) a manuscript is more likely to be cited if it is published in a reputable journal rather than in a journal with a poor reputation (see also Lozano, Larivière, & Gingras, 2012; van Raan, 2012). We include the Journal Impact Factor (JIF) as a measure of the reputation of a journal here. The JIF is a quotient from the sum of citations for a journal in one year and the publications in this journal in the previous two years (Garfield, 2006).

In addition to the three factors that influence citation impact discussed above, we include three more variables. Although the influence of these variables is intended to be reduced with the use of percentiles (a field and age normalised citation impact value where the document type is also controlled), we want to test in this study whether they nevertheless have an impact on the result. (3) First of the three variables is the subject area: The main categories of the Organisation for Economic Co-operation and Development (2007; OECD) are used as a subject area scheme for this study. The OECD scheme provides six broad subject categories for WoS data: (i) Natural Sciences, (ii) Engineering and Technology, (iii) Medical and Health Sciences, (iv) Agricultural Sciences, (v) Social Sciences, and (vi) Humanities. As the numbers in Table 1 show, the publications of the four universities belong to only three subject areas: (i) Natural Sciences, (ii) Engineering and Technology, and (iii) Medical and Health Sciences[1].

(4) The document types included in the study are articles, notes, proceedings papers (published in journals) and reviews. Reviews are usually cited more often than research papers, as they summarise the status of a research subject or area. Since articles as a rule have more research results than notes, we expect that they will have a higher citation impact. Proceedings papers will probably turn out to be less common than highly cited publications as these papers are very often published in an identical form as articles. (5) The final independent variable included in the regression model is the publication year (coded in reverse order so that higher values indicate an older publication, so that 1 = 2010 and 31 = 1980). Regarding this variable, we expect that the opportunity for publications to be cited very frequently increases over time.[2]

The reason for including these variables in this study is not primarily in order to answer content-related questions (such as the extent of the influence of certain factors on citation impact). Regarding some factors influencing citation impact, other more suitable variables have already been proposed: Bornmann, et al. (2011) use, for example, the Normalized Journal Position (NJP) instead of the JIF, with which the importance of a journal can be determined within its subject area – which is not the case with the JIF. The JIF does not offer this subject normalisation but it is specified for each publication in InCites, unlike the NJP. We would like to use the variables included to show the way in which the substantive and practical significant of findings can be determined in addition to statistical significance.

---

[1] Only a few dozen articles were from other fields of study. They were deleted from the analysis.
[2] Table 1 also makes clear that there is tremendous variability across publications in their number of authors and in their length. While the average publication has 4.2 authors, the number of authors across publications ranges between 1 and 23. Even more extreme, while the average publication is only 7.7 pages long, the publications vary anywhere between 1 page and 160 pages in length. In our later analyses we will primarily focus on comparing universities across the ranges of values that tend to occur in practice, but we will also note the implications of our models for publications with more extreme values.

**2.2   Software**

The statistical software package Stata 12 (http://www.stata.com/) is used in this analysis; in particular, we make heavy use of the Stata commands logit, margins, and marginsplot.

**2.3   Analytic Strategy**

To identify citation impact differences between the four universities, we begin by estimating a series of multivariate logistic regression models (Hardin & Hilbe, 2012; Hosmer & Lemeshow, 2000; Mitchell, 2012). Such models are appropriate for the analysis of dichotomous (or binary) responses. Dichotomous responses arise when the outcome is the presence or absence of an event (Rabe-Hesketh & Everitt, 2004). In this study, the binary response is coded as 1 for $P_{top\ 10\%}$ (the document is among the top 10% in citations of all documents) and as 0 otherwise. We then show how various types of Adjusted Predictions and Marginal Effects can make the results for both discrete and continuous variables far more easy to understand and interpret.

# 3   Results

*Logistic Regression models*

Table 2 shows the results for the baseline regression model (model 1) which includes only the universities (and no other variables). As the results show, univ 2, univ 3 and univ 4 have significant fewer highly cited publications than does univ 1 (the reference category). Model 2 includes the possible variables of influence on citation impact in addition to the university variable. It is interesting to see that the differences between universities change substantially with the inclusion of the additional variables. Univ 2 and univ 4 no longer differ significantly from univ 1, while univ 3 performs statistically significantly better than univ 1. This result indicates the importance of taking account of factors that influence citation impact in evaluation studies. Additional analyses (not shown) suggest that this change in position is primarily due to controlling for journal impact. Univ 3 has the lowest average JIF (3.2) while univ 1 has the highest (8.4). Hence, univ 3 "overachieves" in the sense that it gets more citations than can be accounted for by the reputation of journals it publishes in.

*Table 2 about here*

The following results are obtained regarding these factors: (1) publications in Engineering and Technology are more frequently highly cited than publications in other fields (although the difference between Engineering and Technology and Medical and Health Sciences is statistically not significant). This result is counter to expectations and is due presumably to the use of an indicator in this study which is already normalised for the field. (2) Proceedings papers are statistically significantly less likely to be highly cited than other document types. However, differences in the effects of other types of documents are not statistically significant. (3) Publications that were published in journals with a high JIF, that were published longer ago, that have more co-authors, and that are longer in length tend be highly cited more often.

While model 2 fits much better than Model 1, it also makes some questionable assumptions. For example, it assumes that the more pages a paper has, the better. It is probably more reasonable to assume that, after a certain point, additional pages produce less and less benefit or even decrease the likelihood of the paper being cited. Similarly, we might expect diminishing returns for higher JIFs, i.e. it is better to be published in a more influential journal but after a certain point the benefits become smaller and smaller. To address such possibilities, Model 3 adds squared terms for JIF and paper length. Squared terms allow for the possibility that the variables involved eventually have diminishing benefits or even a negative effect on citations, e.g. while a one page paper may be too short to have much impact, a paper that gets too long may be less likely to be read and cited. Both squared terms are negative, highly significant, and theoretically plausible, so we will use Model 3 for the remainder of our analysis.

*Average adjusted predictions (AAPs) and average marginal effects (AMEs) for discrete independent variables*

The logistic regression models illustrate which effects are statistically significant, and what the direction of the effects is, but they give us little practical feel for the substantive significance of the findings. For example, we know that universities differ in their likelihood of being highly cited, but we don't have a practical feeling for how big those differences are. We also know that papers in journals with a higher JIF are more likely to be cited than papers in journals with a lower JIF, but how much more likely? The addition of squared terms makes interpretation even more difficult. Adjusted predictions and marginal effects can provide clearer pictures of these issues. First, we will present the adjusted predictions and marginal effects, and then we will explain how those values can be computed for discrete variables.

*Table 3 about here*

The first column of Table 3 shows the average adjusted predictions (AAPs) for the discrete variables in the final logistic regression model, while the second column displays their Average Marginal Effects (AMEs). The two columns are very helpful in clarifying the magnitudes of the effects of the different independent variables. The AAPs in column 1 show that – after other variables are taken into account – about 16.2% of univ 1's publications are highly cited, compared to almost 24.5% of univ 3's. The AMEs in column 2 show how the AAPs for each category differ from that of the reference category. So, the AME of .0829 for univ 3 means that 8.3% more of univ 3's publications are highly cited than are univ 1's (i.e. 24.5% - 16.2% = 8.3%). Again, remember that this is after controlling for other variables. For whatever reason, univ 3's papers are more likely to be highly cited than would be expected based on their values on the other variables in the model. This might reflect, for example, that univ 3 tends to publish more on topics that are of broader interest even though they appear in journals with a lesser impact overall. Whatever the reasons for the difference, the adjusted predictions and the marginal effects probably provide a much clearer picture of the differences across universities than the logistic regressions did.

Similarly, we see that – after controlling for other variables – more than a quarter (26.5%) of the publications in Engineering and Technology are highly cited, compared to a little over a fifth of those in the Medical and Health Sciences (22.3%) The AMEs in Column 2 of Table 3 shows that this difference of 4.28% is statically significant. In other words, even after controlling for all the other variables in the model, 4.3% more of Engineering and Technology papers are highly cited than is the case for papers in the Medical and Health Sciences. The AAPs and the AMEs further show us that Engineering and Technology papers also have an advantage of about 6.8% over papers in the Natural Sciences. Again, the coefficients from the logistic regressions had already shown us that papers in Engineering and Technology were more likely to be highly cited than papers in other fields, but the AAPs and AMEs give us a much more tangible feel for just how much more likely.

Table 3 further shows us that, after adjusting for the other variables in the model, 20.8% of articles, 22.2% of notes, 15.7% of proceedings papers, and 24.4% of reviews are highly cited. The marginal effects show that the differences between articles and proceedings papers is statistical significant, while the difference between articles and reviews falls just short of statistical significance.

Examining exactly how the AAPs and AMEs are computed for categorical variables will help to explain the approach. For convenience, we will focus on the university variable, but the logic is the same for document type and subject area. Intuitively, the AAPs and the AMEs for the universities are computed as follows:

- Go to the first publication. Treat that publication as though it were from univ 1, regardless of where it actually came from. Leave all other independent variable values as is. Compute the probability that this publication (if it were from univ 1) would be highly cited. We will call this AP1 (where 1 refers to the category of the independent variable that we are referring to, i.e. the predicted probability of $P_{top\ 10\%}$ which this publication would have if it came from univ 1).

- Now do the same for each of the other universities, e.g. treat the publication as though it was from univ 2, univ 3, or univ 4, while leaving the other variables at their observed values. Call the predicted probabilities AP2 through AP4.

- Differences between the computed probabilities give you the marginal effects for that publication, i.e., ME2 = AP2 − AP1, ME3 = AP3 − AP1, ME4 = AP4 − AP1.

- Repeat the procedure for every case in the sample.

- Compute the averages of all the individual adjusted predictions you have generated. This will give you AAP1 through AAP4. Similarly, compute the averages of the individual marginal effects. This gives you AME2 through AME4.

With AAPs and AMEs for discrete variables, in effect different hypothetical populations are compared – one where every publication is from univ 1, another where every publication is from univ 2, etc. – that have the exact same values on the other independent variables in the regression model. The logic is similar to that of a matching study, where subjects have identical values on every independent variable except one (Williams, 2012).

Since the only difference between these publication populations is their university (their origin), the university must be the cause of the differences in their probability of being highly cited[3].

*Average adjusted predictions (AAPs) and average marginal effects (AMEs) for continuous independent variables*

The effects of continuous variables (e.g. the JIF) in a logistic regression model are, other than their sign and statistical significance, also difficult to interpret. For example, publications in journals with high JIFs tend to be more frequently highly cited than publications in journals with low JIFs. The question is: How much more often is that the case? Continuous variables offer additional challenges in that (a) they have many more possible values than do discrete variables – indeed a continuous variable can potentially have an infinite number of values – and (b) the calculation of marginal effects is different for continuous variables than it is for discrete variables. It is therefore difficult (or, at least, of limited value) to come up with a single number that represents any sort of "average" effect for a continuous variable. Instead, it is useful to compute the Average Adjusted Predictions (AAPs) and Average Marginal Effects (AMEs) across a range of the variable's plausible (or at least possible) values.

*Figure 1 about here*

Figure 1 therefore presents the AAPs for JIF. The grey bands represent the 95% confidence interval for each predicted value. AAPs are estimated for JIF values ranging between 0 and 35. We chose an upper bound of 35 because less than 1% of all publications have a higher JIF than that.

---

[3] Another popular way of getting at the idea of "average" values uses Adjusted Predictions at the Means (APMs) and Marginal Effects at the Means (MEMs). With this approach, rather than use all of the observed values for all the publications, the mean values for each independent variable are computed and then used in the calculations. While widely used, this approach has various conceptual problems, e.g., a publication cannot be .5 of univ 1 or .1 of univ 2. In our examples, the means approach produces similar results to those presented here, but that is not always the case.

The figure shows that, not surprisingly, publications in journals with higher JIFs are more likely to be highly cited than publications in journals with low JIFs. We already knew that from the logistic regressions, but plotting the AAPs makes it much clearer how great the differences are. Publications with a JIF of close to 0 have less than a 10% chance of being highly cited. A publication with a JIF of 10, however, has almost a 48% predicted probability of being highly cited. (Only about 8% of all publications have a JIF of 10 or higher, which means that publications that have a JIF of 10 are appearing in some of the most influential journals.) Publications in the most elite journals with a JIF of 30 have about an 88% predicted probability of being highly cited.

The graph also reveals, however, that the beneficial effect of higher JIFs gradually decline as the JIF gets higher and higher. That is, the curve depicting the JIF predictions gradually becomes less and less steep. While there is a big gain in going from a JIF of 0 to 10, there is virtually no gain in going from a JIF of 25 to a JIF of 35. As we speculated earlier, after reaching a certain point there is little or nothing to be gained from publishing in a journal that has an ever higher JIF.

*Figure 2 about here*

The AMEs for JIF that are presented in Figure 2 further illustrate the declining benefits to higher JIFs. Initially, changes in JIFs between 0 and 10 produce greater and greater increases in the likelihood of being highly cited. For example, going from a JIF of 0 to a JIF of 1 produces some increase in the likelihood of being highly cited, but going from 9 to 10 produces an even greater benefit. For JIFs between 10 and 30, however, additional increases in JIFs produce smaller (but still positive) increases in the likelihood of being highly cited.

After the JIF hits 30, though, there are no additional benefits to being in a journal that has an even higher JIF[4].

*Figures 3 and 4 about here*

Figures 3 and 4 present similar analyses. Figure 3 presents the AAPs for document length, for values ranging between 1 page and 120 pages. This is a very wide range – 99% of all documents are 25 pages are less – but it illustrates the estimated declining benefits as papers get longer and longer.

As Figure 3 shows, a 1 page paper has only about a 14% predicted probability of being highly cited, while an average length paper (about 8 pages) has an AAP of almost 21%. However, the benefits of greater length gradually become smaller and smaller. While an 80 page paper has an 80% predicted probability of being highly cited, making a publication longer than that actually reduces the likelihood of it being highly cited.

The AMEs for document length presented in Figure 4 further clarify the at first rising and then declining effects of increases in document length. Up until about 20 pages, the benefits of greater document length get greater and greater, i.e. while moving from 1 page to 2 is good, moving from 19 pages to 20 is even better. But, after 20 pages, the benefits of greater document length get smaller and smaller, and by about 80 pages (85 to be precise) any additional pages actually reduce the likelihood of being highly cited. Of course, given how few documents approach such lengths, and given the huge confidence intervals for the estimates, we should view such conclusions with some caution.

*Adjusted predictions at representative values (APRs) and marginal effects at representative values (MERs) for continuous and discrete variables together*

---

[4] Indeed, if we extend the graphs to include even higher values of JIF, gains in JIF actually produce declines in the likelihood of being highly cited, e.g. it is better to have a JIF of 30 than it is to have a JIF of 50. This is a necessary consequence of including squared terms in the model. In practice, however, hardly any publications have JIFs higher than 35. We should be careful about making predictions involving values that generally fall well outside most of the observed values in the data.

As we show with our example of four universities, the AAPs and AMEs provide a much clearer feel for the differences that exist across categories or ranges of the independent variables than statistical significance testing can. Still, as Williams (2012) points out, the use of averages with discrete variables can obscure important differences across publications. In reality, the effect that variables like universities, document type, and subject area have on the probability of being highly cited need not be the same for every publication. For example, as Williams (2012) shows in his analysis of data from the early 1980s, racial differences in the probability of diabetes are very small at young ages. This is primarily because young people, white or black, are very unlikely to have diabetes. As people get older, the likelihood of diabetes gets greater and greater; but it goes up more for blacks than it does for whites, hence racial differences in diabetes are substantial at older ages.

In the case of the present study, Table 3 showed us that, on average, publications from univ 3 were about 8.3 percentage points more likely to be highly cited than publications from univ 1. But, this gap almost certainly differs across values of the other independent variables. For example, a 1 page paper, or a paper with a low JIF, isn't that likely to be highly cited regardless of which university it came from. But, as increases in other variables increase the likelihood of a publication being highly cited, the differences in the adjusted predictions across universities will likely increase as well.

Williams (2012) therefore argues for the use of marginal effects at representative values (MERs) and, by logical extension, adjusted predictions at representative values (APRs). These approaches basically combine analysis of the effects of discrete and continuous variables simultaneously. With APRs and MERs, plausible or at least possible ranges of values for one or more continuous independent variables are chosen. We then see how the adjusted predictions and marginal effects for discrete variables vary across that range.

*Figures 5 and 6 about here*

Figure 5 shows the APRs for the four universities for JIFs ranging between 0 and 13. Thirteen is chosen because 95% of all publications have JIFs of 13 or less; extending the range to include larger values than 13 makes the graph harder and harder to read. The graph shows that, for all four universities, increases in JIFs increase the likelihood of the publication being highly cited. But, for JIFs near 0, the differences between univ 3 and the others are small – about a 4 percentage point difference. However, as the JIFs increase, the gap between univ 3 and the others becomes greater and greater. When the JIF reaches 13, univ 3 has about 14 percentage points more of its publications highly cited than do the others. Figure 6, which shows the MERs, makes it even clearer that a fairly small gap between the universities at low JIFs gets much larger as the JIF gets bigger and bigger.

*Figures 7 and 8 about here*

Figures 7 and 8 show the APRs and MERs for the four universities across varying document lengths. About 99% of all papers are 25 pages or less so we limit the range accordingly. Again, for all four universities, the longer the document is, the higher the predicted probability is that it will be highly cited. However, for a 1 page paper, the predicted difference between univ 3 and the other universities is only about 6%. But, for a 25 page paper, the predicted gap is much larger, about 13%. The MERs presented in Figure 8 are another way of showing how the predicted gap between universities gets greater and greater as the page length gets longer and longer.

## 4    Discussion

When we compare research institutions in evaluative bibliometrics we are primarily interested in the differences that are significant in practical terms. Statistical significance tests in this context only provide information on whether an effect that has been determined in a random sample applies beyond the random sample. These tests do not however indicate how large the effect is (Schneider, in press) nor whether differences have a practical significance (Williams, 2012). One way to reveal significant differences is to work with Goldstein-adjusted confidence intervals (Bornmann, Mutz, & Daniel, in press). With these confidence intervals, it is possible to interpret the significance of differences among research institutions meaningfully: For example, rank differences in the Leiden Ranking among universities should be interpreted as meaningful only if their confidence intervals do not overlap.

In this paper we present a different approach, and one which can be easily adapted to a wide array of substantive topics. With techniques like logistic regression, it is easy to determine the direction of effects and their statistical significance, but it is far more difficult to get a practical feel for what the effects really mean. In the present example, the logistic regressions showed us that, after controlling for other variables, univ 3 was more likely to have its publications highly cited than were other universities. We should be careful about interpreting this as meaning that univ 3 is "better" than its counterparts; for example, besides being highly cited, we might expect a good university to place more of its papers in high impact journals, and univ 3 actually fares the worst in this respect. But the results do mean that, for whatever reason, univ 3 is more likely to have its publications highly cited than would be expected on the basis of its values on the other variables considered by the model. Further research might yield insights into what exactly univ 3 is doing that make its publications disproportionately successful.

The logistic regression results also make clear that, for example, longer papers (at least up to a point) get cited more than shorter papers and publications in high impact journals get cited more than publications in low impact journals. The logistic regression results fail to make clear, however, how large and important these effects are in practice. The use of average adjusted predictions (AAPs) and average marginal effects (AMEs) – along with average predictions at representative values (APRs) and marginal effects at representative values (MERs) – helped make these effects much more tangible and easier to grasp. We saw, for example, that, after controlling for other variables, on average univ 3 had about 8 percentage points more of its publications highly cited than did other universities. But, the expected gap was much smaller for very short documents and documents in low impact journals (which, regardless of which university they come from, tend not to be heavily cited). Conversely the gap between the universities was much greater for longer papers and higher impact journals. The magnitudes of other effects, such as subject area and document type, were also made explicit.

The analyses yielded a number of other interesting insights. They illustrated, for exampling, the diminishing and even negative returns as papers got longer and longer. They suggested that, after a certain point (about 25) higher JIFs produced little or no additional benefits.

We hope that with this paper we are making a contribution to enabling the measurement of not only statistical significance but also practical significance in evaluative bibliometric studies. These studies would then comply with the publication guidelines such as those of the American Psychological Association (2009) which recommend both significance and substantive tests for empirical studies. Effect size is crucial particularly in evaluative bibliometrics, as far-reaching decisions on careers and financing are often made on the basis of publication and citation data. The effect size gives information about how well a research institution is performing compared to another. Bornmann (in press) has already presented a number of tests for effective size measurement. The use of adjusted predictions and marginal effects provide alternative ways by which differences across institutions can be visualized and made easier to interpret.

Table 1.

Description of the dependent and independent variables (*n*=15,426 publications)

| Variable | Percentage / Mean | Standard deviation | Minimum | Maximum |
| --- | --- | --- | --- | --- |
| **Dependent variable:** | | | | |
| $PP_{top\ 10\%}$ | 20.7% | | 0 | 1 |
| **Independent variable:** | | | | |
| University | | | | |
|   Univ 1 (reference category) | 7.4% | | 0 | 1 |
|   Univ 2 | 3.3% | | 0 | 1 |
|   Univ 3 | 55.4% | | 0 | 1 |
|   Univ 4 | 33.9% | | 0 | 1 |
| Subject area | | | | |
|   Engineering and Technology (reference category) | 11.4% | | 0 | 1 |
|   Medical and Health Sciences | 10.7% | | 0 | 1 |
|   Natural Sciences | 77.9% | | 0 | 1 |
| Document type | | | | |
|   Article (reference category) | 82.9% | | 0 | 1 |
|   Note | 4.3% | | 0 | 1 |
|   Proceedings Paper | 9.7% | | 0 | 1 |
|   Review | 3.2% | | 0 | 1 |
| Journal Impact Factor | 4.5 | 5.8 | 0.4 | 54.3 |
| Years since Publication (1=2010, 31=1980) | 17.7 | 8 | 1 | 31 |
| Number of Authors | 4.2 | 2.4 | 1 | 23 |
| Number of Pages | 7.7 | 6.1 | 1 | 160 |

Table 2. Logistic Regression Models for PP$_{top\ 10\%}$

|  | (1) Baseline | (2) All variables | (3) Squared terms added |
|---|---|---|---|
| *University* | | | |
| Univ 2 | -0.716*** | -0.184 | 0.0245 |
|  | (-5.16) | (-1.12) | (0.15) |
| Univ 3 | -0.541*** | 0.375*** | 0.640*** |
|  | (-7.51) | (4.19) | (7.06) |
| Univ 4 | -0.195** | 0.0989 | 0.135 |
|  | (-2.64) | (1.13) | (1.55) |
| *Subject Area* | | | |
| Medical and Health Sciences |  | -0.162 | -0.280** |
|  |  | (-1.62) | (-2.74) |
| Natural Sciences |  | -0.342*** | -0.464*** |
|  |  | (-4.89) | (-6.48) |
| *Document Type* | | | |
| Note |  | 0.0589 | 0.0963 |
|  |  | (0.54) | (0.86) |
| Proceedings Paper |  | -0.614*** | -0.410*** |
|  |  | (-6.14) | (-4.03) |
| Review |  | 0.233 | 0.241 |
|  |  | (1.90) | (1.96) |
| *Further variables* | | | |
| Journal Impact Factor |  | 0.149*** | 0.308*** |
|  |  | (27.81) | (30.28) |
| Years Since Publication |  | 0.0259*** | 0.0328*** |
|  |  | (8.73) | (10.81) |
| Number of Authors |  | 0.0626*** | 0.0511*** |
|  |  | (6.55) | (5.27) |
| Number of Pages |  | 0.0600*** | 0.0878*** |
|  |  | (13.42) | (14.53) |
| Journal Impact Factor Squared |  |  | -0.00502*** |
|  |  |  | (-19.44) |
| # of Pages Squared |  |  | -0.000519*** |
|  |  |  | (-6.86) |
| _cons | -0.968*** | -3.124*** | -3.961*** |
|  | (-14.61) | (-23.51) | (-27.53) |
| N | 15426 | 15426 | 15426 |
| pseudo $R^2$ | 0.007 | 0.126 | 0.148 |
| AIC | 15617.5 | 13763.8 | 13419.5 |
| BIC | 15648.1 | 13863.2 | 13534.2 |
| chi2 | 104.3 | 1976.0 | 2324.3 |
| D.F. | 3 | 12 | 14 |

Notes.
$z$ statistics in parentheses
* $p < 0.05$, ** $p < 0.01$, *** $p < 0.001$

Table 3. Average adjusted predictions (AAPs) and average marginal effects (AMEs) for the discrete variables in the regression model (n=15,426 publications)

|  | (1) AAPS | (2) AMES |
|---|---|---|
| *University* | | |
| Univ 1 | 0.162*** | |
|  | (18.52) | |
| Univ 2 | 0.164*** | 0.00270 |
|  | (10.08) | (0.15) |
| Univ 3 | 0.245*** | 0.0829*** |
|  | (48.65) | (7.97) |
| Univ 4 | 0.177*** | 0.0154 |
|  | (39.27) | (1.59) |
| *Subject Area* | | |
| Engineering and Technology | 0.265*** | |
|  | (24.69) | |
| Medical and Health Sciences | 0.223*** | -0.0428** |
|  | (21.06) | (-2.76) |
| Natural Sciences | 0.197*** | -0.0679*** |
|  | (59.81) | (-6.05) |
| *Document Type* | | |
| Article | 0.208*** | |
|  | (64.00) | |
| Note | 0.222*** | 0.0136 |
|  | (14.04) | (0.84) |
| Proceedings paper | 0.157*** | -0.0509*** |
|  | (14.34) | (-4.42) |
| Review | 0.244*** | 0.0352 |
|  | (13.03) | (1.86) |

Notes.
$z$ statistics in parentheses
$^{*}$ $p < 0.05$, $^{**}$ $p < 0.01$, $^{***}$ $p < 0.001$

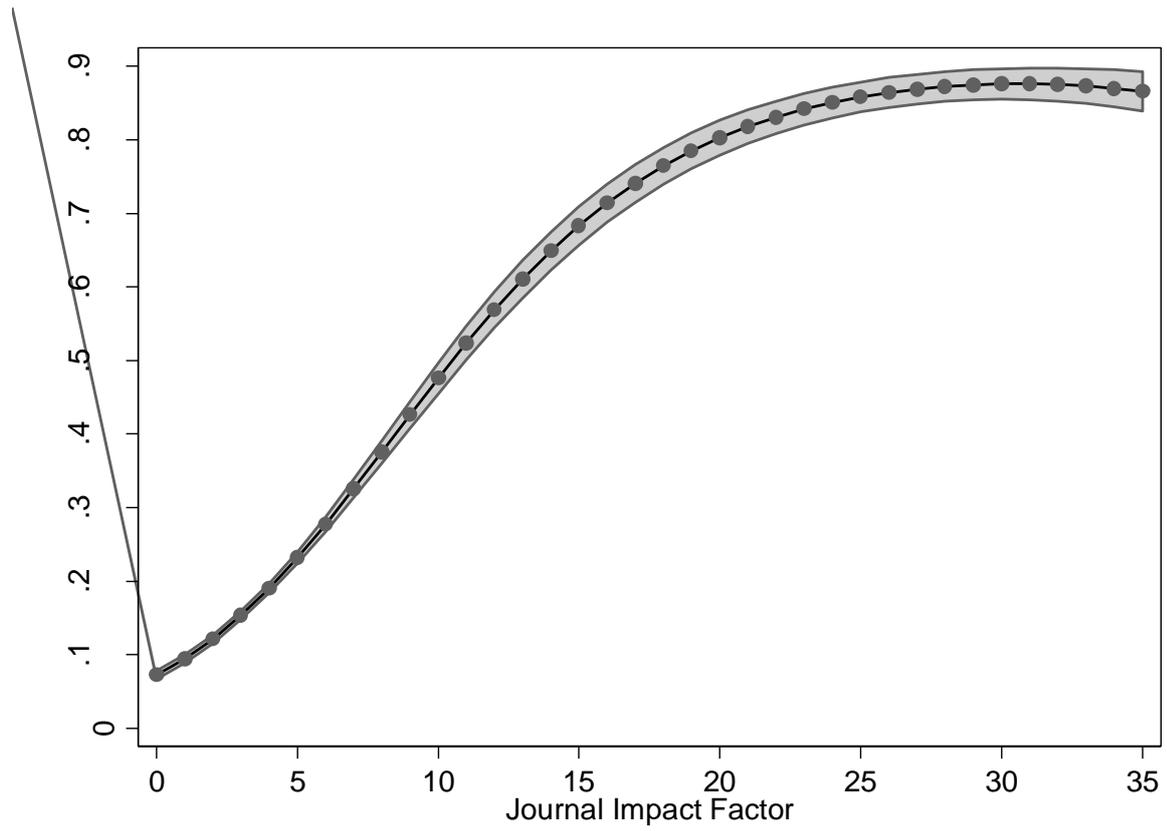

Figure 1. Average Adjusted Predictions & 95% Confidence Intervals for Journal Impact Factor.

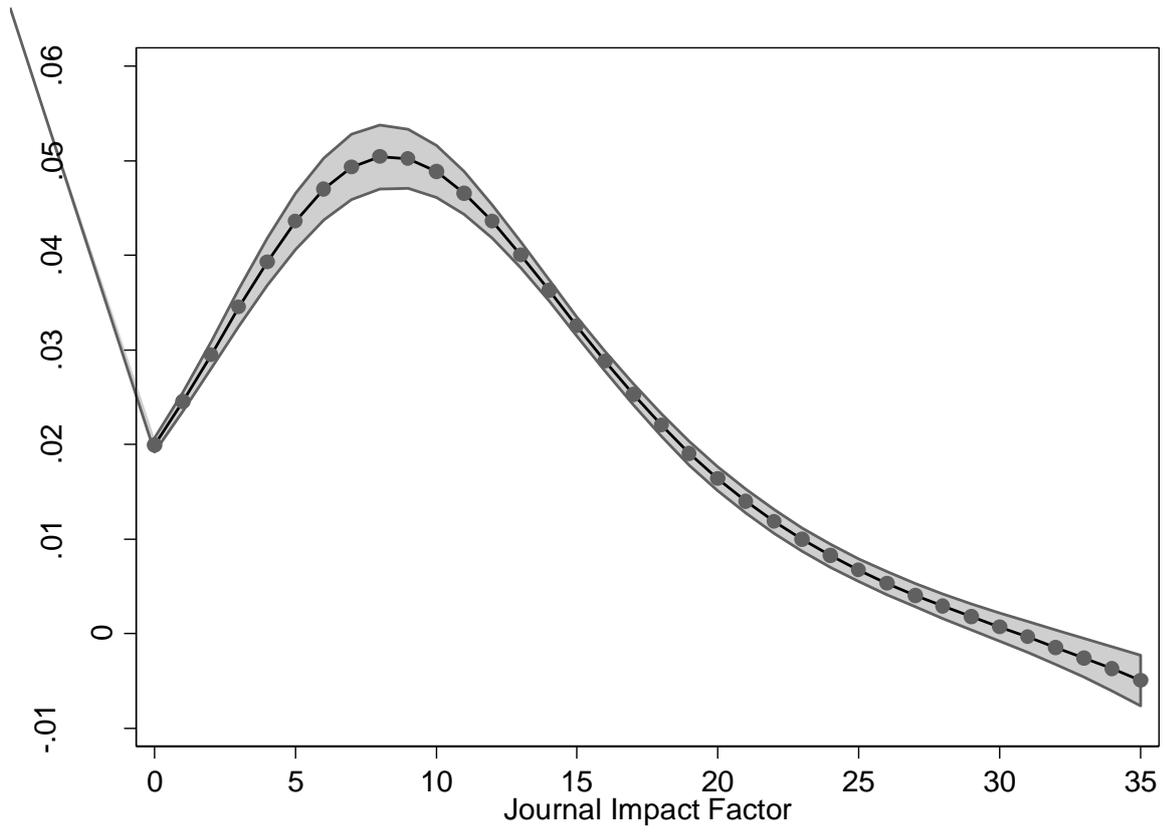

Figure 2. Average Marginal Effects & 95% Confidence Intervals for Journal Impact Factor.

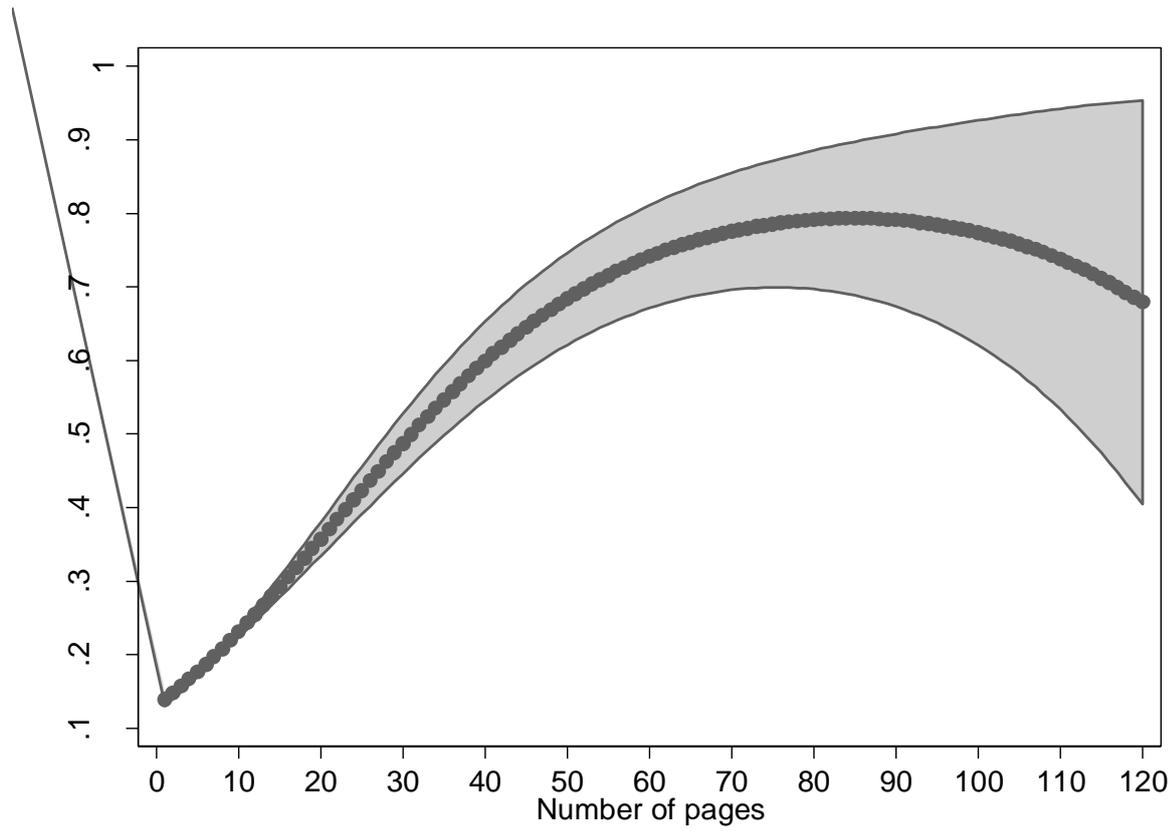

Figure 3. Average Adjusted Predictions & 95% Confidence Intervals for Length of Document.

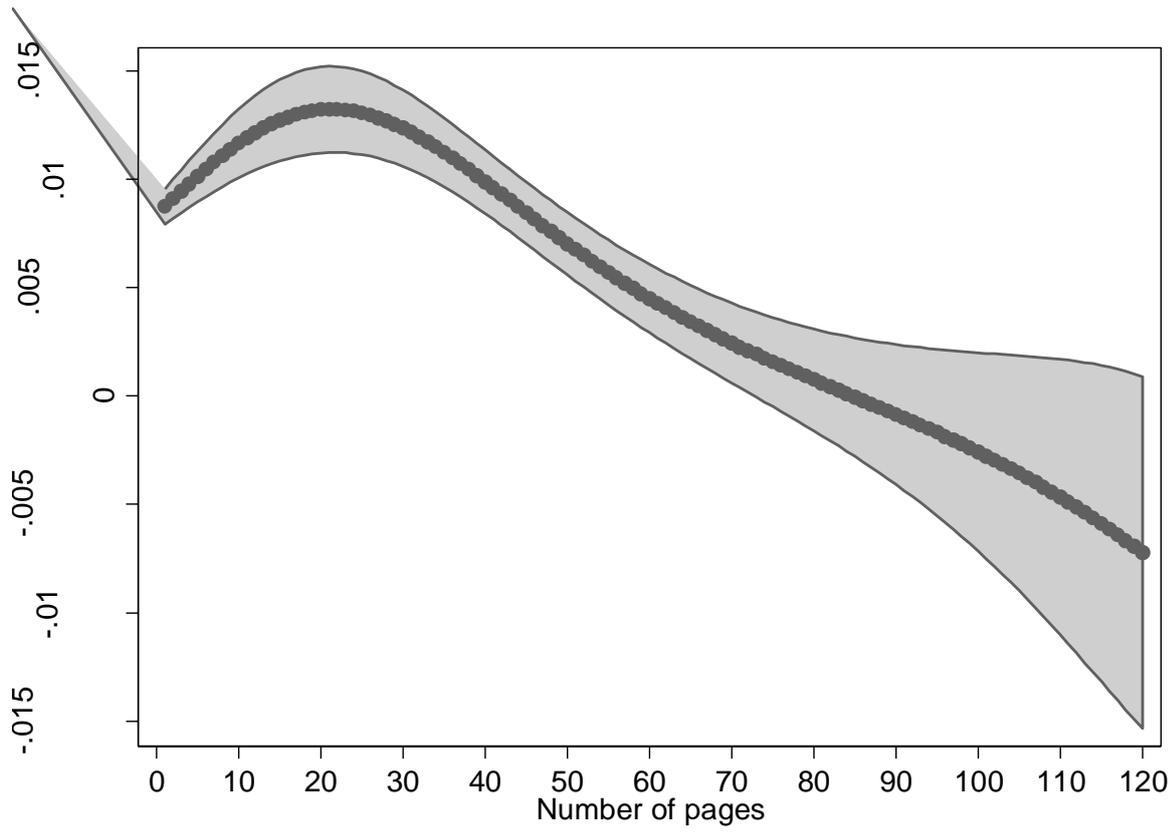

Figure 4. Average Marginal Effects & 95% Confidence Intervals for Length of Document.

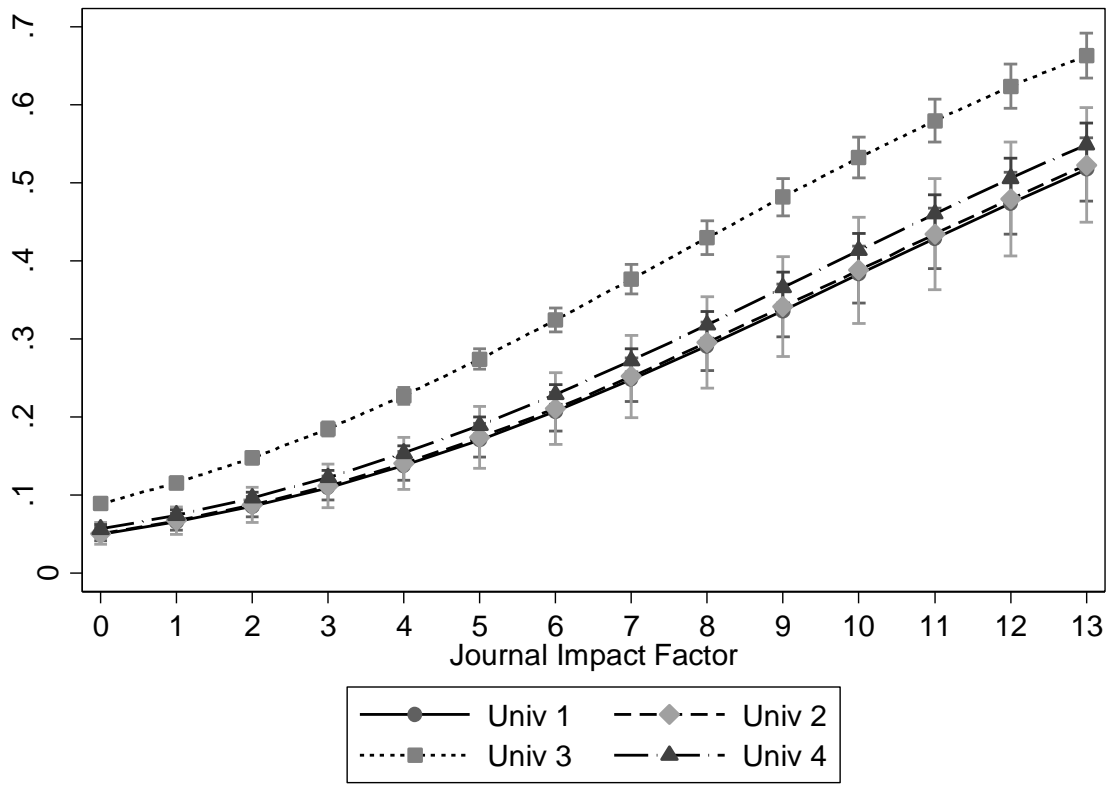

Figure 5. Adjusted Predictions at Representative Values & 95% Confidence Intervals for Four Universities and Journal Impact Factor.

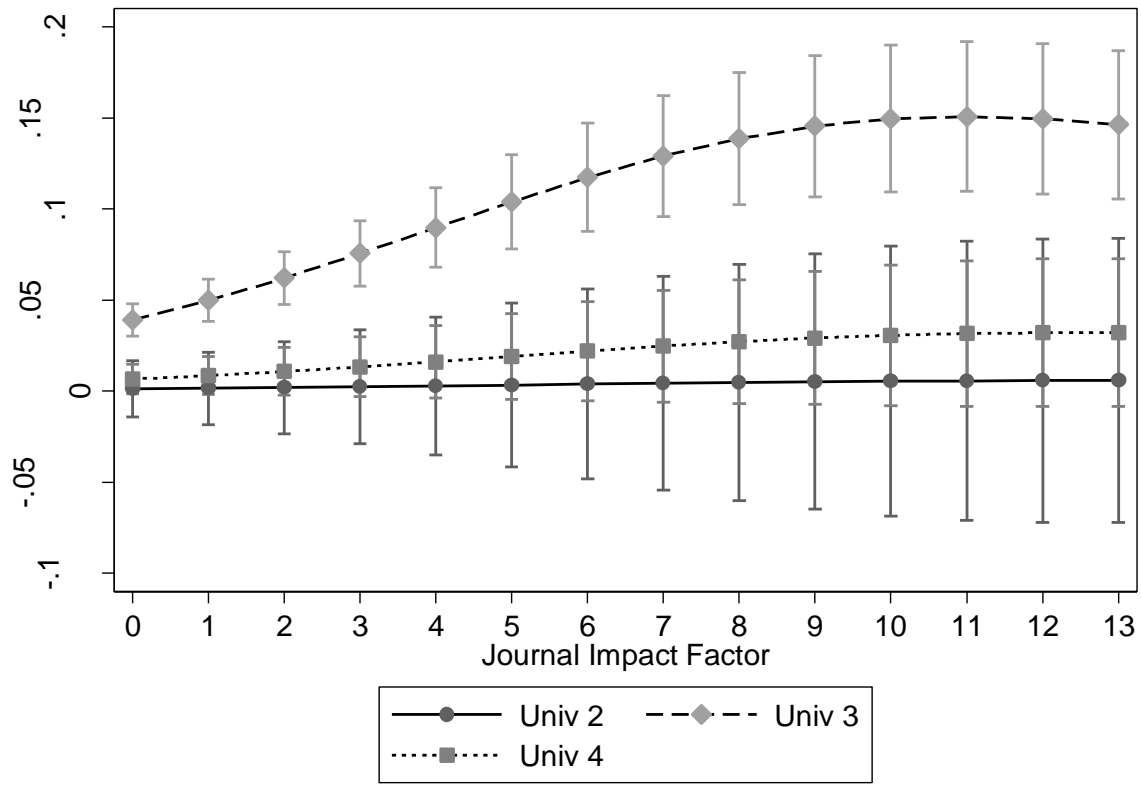

Figure 6. Marginal Effects at Representative Values & 95% Confidence Intervals for Four Universities and Journal Impact Factor.

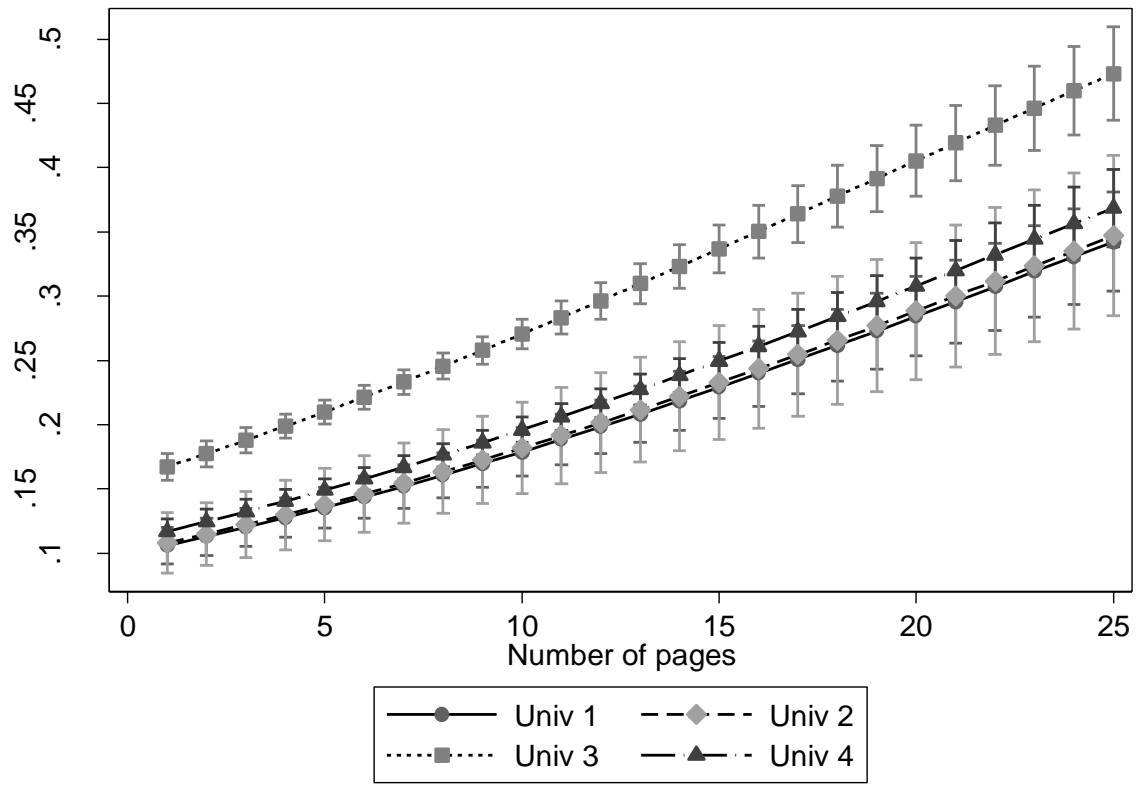

Figure 7. Adjusted Predictions at Representative Values & 95% Confidence Intervals for Four Universities and Document Length.

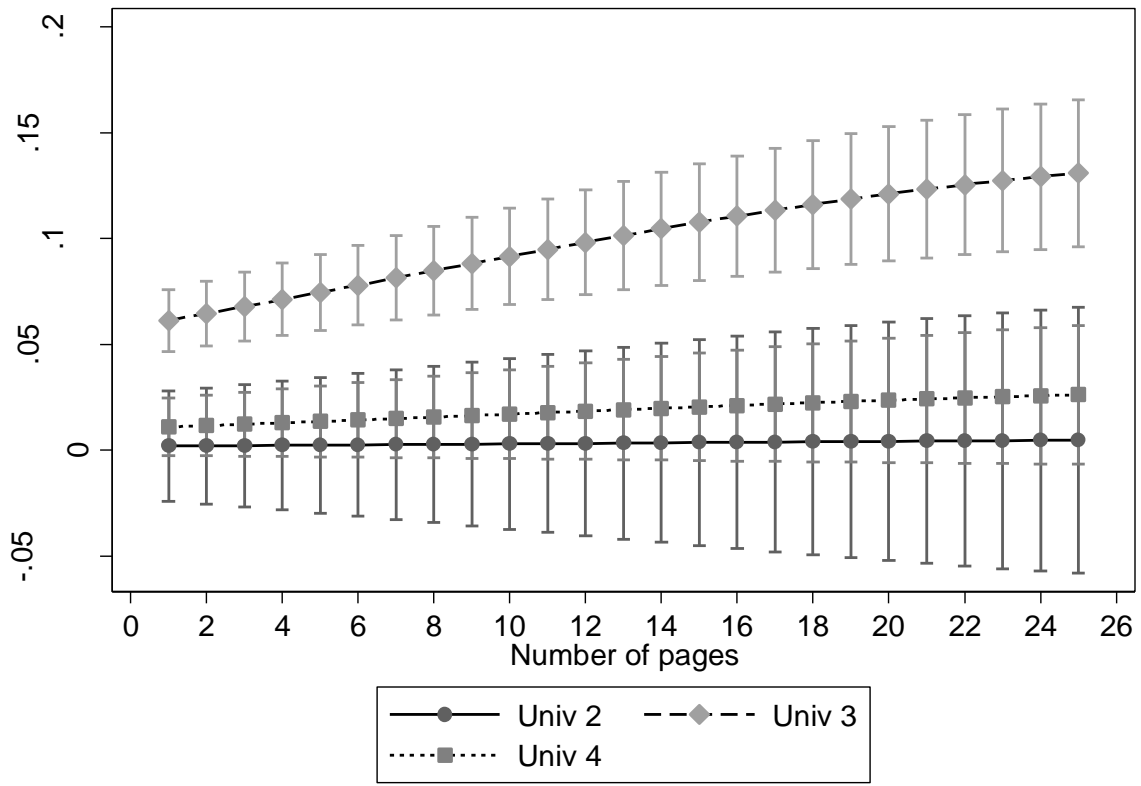

Figure 8. Marginal Effects at Representative Values & 95% Confidence Intervals for Four Universities and Document Length.